\documentclass[12pt,draft]{article} %fleqn
\usepackage{amsmath}
\usepackage{amssymb}
\usepackage{dsfont}
\usepackage{amsthm}
   \font\twelvebm                       = cmmib10 at 12truept
   \font\tenbm                          = cmmib10 at 10truept
   \font\sevenbm                        = cmmib10 at 7truept
  at 10truept  at 10truept  at
10truept  at 10truept
 at 10truept
\mathchardef \BGamma            = "0900 \mathchardef \BDelta =
"0901 \mathchardef \BTheta            = "0902 \mathchardef
\BLambda           = "0903 \mathchardef \BXi               = "0904
\mathchardef \BPi               = "0905 \mathchardef \BSigma =
"0906 \mathchardef \BUpsilon          = "0907 \mathchardef \BPhi =
"0908 \mathchardef \BPsi              = "0909 \mathchardef \BOmega
= "090A \mathchardef \Balpha            = "090B \mathchardef
\Bbeta             = "090C \mathchardef \Bgamma = "090D
\mathchardef \Bdelta            = "090E \mathchardef \Bepsilon =
"090F \mathchardef \Bzeta             = "0910 \mathchardef \Beta =
"0911 \mathchardef \Btheta = "0912 \mathchardef \Biota = "0913
\mathchardef \Bkappa            = "0914 \mathchardef \Blambda
= "0915 \mathchardef \Bmu               = "0916 \mathchardef \Bnu
= "0917 \mathchardef \Bxi               = "0918 \mathchardef \Bpi
= "0919 \mathchardef \Brho              = "091A \mathchardef
\Bsigma            = "091B \mathchardef \Btau = "091C \mathchardef
\Bupsilon          = "091D \mathchardef \Bphi = "091E \mathchardef
\Bchi              = "091F \mathchardef \Bpsi = "0920 \mathchardef
\Bomega            = "0921 \mathchardef \Bvarepsilon       = "0922
\mathchardef \Bvartheta         = "0923 \mathchardef \Bvarpi
= "0924 \mathchardef \Bvarrho = "0925 \mathchardef \Bvarsigma
= "0926 \mathchardef \Bvarphi           = "0927
\mathchardef \bA        = "0941 \mathchardef \bB        = "0942
\mathchardef \bC        = "0943 \mathchardef \bD        = "0944
\mathchardef \bE        = "0945 \mathchardef \bF        = "0946
\mathchardef \bG        = "0947 \mathchardef \bH        = "0948
\mathchardef \bI        = "0949 \mathchardef \bJ        = "094A
\mathchardef \bK        = "094B \mathchardef \bL        = "094C
\mathchardef \bM        = "094D \mathchardef \bN        = "094E
\mathchardef \bO        = "094F \mathchardef \bP        = "0950
\mathchardef \bQ        = "0951 \mathchardef \bR        = "0952
\mathchardef \bS        = "0953 \mathchardef \bT        = "0954
\mathchardef \bU        = "0955 \mathchardef \bV        = "0956
\mathchardef \bW        = "0957 \mathchardef \bX        = "0958
\mathchardef \bY        = "0959 \mathchardef \bZ        = "095A
\mathchardef \ba        = "0961 \mathchardef \bb        = "0962
\mathchardef \bc        = "0963 \mathchardef \bd        = "0964
\mathchardef \bee       = "0965 %%%I CHANGED IT FROM \be; SEE IN THE BEGGINING.
\mathchardef \bff       = "0966 \mathchardef \bg        = "0967
\mathchardef \bh        = "0968
\mathchardef \bj        = "096A \mathchardef \bk        = "096B
\mathchardef \bl        = "096C \mathchardef \bm        = "096D
\mathchardef \bn        = "096E \mathchardef \bo        = "096F
\mathchardef \bp        = "0970 \mathchardef \bq        = "0971
\mathchardef \br        = "0972 \mathchardef \bs        = "0973
\mathchardef \bt        = "0974 \mathchardef \bu        = "0975
\mathchardef \bv        = "0976 \mathchardef \bw        = "0977
\mathchardef \bx        = "0978 \mathchardef \by        = "0979
\mathchardef \bz        = "097A

\font\tencb            = cmssbx10 scaled \magstep4 \font\eigcb =
cmssbx10 scaled \magstep2 \textfont8             = \tencb
\mathchardef\bAs       = "1841
\def\Asem#1#2{\mathop{\vrule height10.5pt depth5.5pt width0pt\bAs}_{#1}^{#2}}
\def\asem#1#2{
          \ifmmode
         \ifinner
            \raise0.9pt\hbox{$\scriptstyle\bAs$}_{#1}^{#2}
         \else
            \Asem{#1}{#2}
         \fi
          \fi
          }

\newtheorem{theo}{\small\bf Theorem}

\newtheorem{rem}{\small\bf Remark}

\newtheorem{defi}{\small\bf Definition}
\newenvironment{DEFI}{\begin{defi} \rm}{\end{defi}}

\renewcommand{\Pr}{\mathds{P}}

\newcommand{\be}{\begin{equation}}
\newcommand{\ee}{\end{equation}}
\newcommand{\E}{\mathds{E}}
\newcommand{\Var}{\mbox{\rm \hspace*{.2ex}Var\hspace*{.2ex}}}
\newcommand{\Cov}{\mbox{\rm \hspace*{.2ex}Cov\hspace*{.2ex}}}

\newcommand{\spam}{\mbox{\rm{spam}}}
\newcommand{\Z}{\mathds{Z}}

\newcommand{\R}{\mathds{R}}

\newcommand{\CB}{\mathcal{B}}
\newcommand{\CH}{\mathcal{H}}

\newcommand{\IP}{\mbox{IP}(\mu;\delta,\beta,\gamma)}
\newcommand{\CO}{\mbox{CO}(\mu;\delta,\beta,\gamma)}

\newcommand{\bbb}[1]{\mbox{\boldmath $ #1 $}}

\newenvironment{pr}[1]{{\small\bf {#1}:}}{}
 \title{ \Large\bf On matrix variance inequalities
 \footnote{Work partially supported by the University of
 Athens' Research fund under Grant 70/4/5637.}\vspace*{-.6em}}

 \author{\large
 G.\ Afendras \ \ and \ \ N.\ Papadatos\footnote{Corresponding
 author. {\tt e-mail address:\ npapadat@math.uoa.gr}}}

 \date{\small
 Department of Mathematics, Section of Statistics and O.R.,
 University of Athens, \\ Panepistemiopolis, 157 84 Athens, Greece.
 }

\begin{document}

 \maketitle \vspace*{-2em}

 \thispagestyle{empty}

 \begin{abstract}
  \noindent
 Olkin and Shepp (2005, {\it J.\ Statist.\ Plann.\ Inference},
 vol.\ 130, pp.\ 351--358)
 presented a matrix form of Chernoff's inequality for
 Normal and Gamma (univariate) distributions. We extend and
 generalize this
 result, proving Poincar\'e-type and Bessel-type inequalities, for matrices
 of arbitrary order
 and for a large class of distributions.
 \end{abstract}
 {\footnotesize {\it MSC}:  Primary 60E15.
 }
 \vspace*{-1em}

 \section{Introduction}
 \vspace*{-.5em}
 \label{sec1}

 %$\boldsymbol{+\infty\int}$
 Let $Z$ be a standard Normal random variable (r.v.)\
 %$Z$
 %be a $\CN(0,1)$ r.v.\
 and assume that $g_1,\ldots,g_p$ are
 absolutely continuous, real-valued, functions of $Z$,
 each with finite variance (with respect to $Z$).
 Olkin and Shepp (2005) presented a matrix extension
 of Chernoff's variance inequality, which reads as
 \vspace{.5em}
 follows:

 \noindent
 {\small\bf Olkin and Sepp (2005).} If $\bbb{D}=\bbb{D}(\bbb{g})$ is
 the covariance matrix of
 the random vector $\bbb{g}=(g_1(Z),\ldots,g_p(Z))^{\mbox{t}}$,
 where `$\mbox{t}$' denotes transpose,
 and if $\E[g'_i(Z)]^2<\infty$
 for all $i=1,2,\ldots,p$, then
 \[
 \bbb{D}\leq \bbb{H},
 \]
 where $\bbb{H}=\bbb{H}(\bbb{g})=(\E[g_i'(Z)g_j'(Z)])_{p\times p}$, and the inequality
 is considered in the sense of Loewner ordering, that is, the matrix
 $\bbb{H}-\bbb{D}$ is nonnegative definite.
 \vspace{.5em}

 %(see  \cite{OL}) proved, expecting any function
 %$g_j$, $j=1,\ldots,p$,
 %in orthonormalized Hermite polynomials, that the matrix $H-D$ is
 %semi-positively defined, where $D$ is a dispersion matrix of
 %functions $g_1(Z),\ldots,g_p(Z)$ and
 %$H=\Big(\E\big[g'_i(Z)g'_j(Z)\big]\Big)$
 %is a $p\times p$ matrix.
 In this note we extend and generalize
 this inequality for a large family of discrete and continuous
 r.v.'s. Specifically, our results apply
 to any r.v.\ $X$ according to
 one of the following definitions
 (c.f.\ \cite{APP1}).
 %\bigskip

 \noindent
 \begin{DEFI}
 {\small\bf (Integrated Pearson Family).}
 Let $X$ be an r.v.\ with probability density function
 (p.d.f.)\ $f$ and finite
 mean $\mu=\E(X)$. We say that $X$ follows
 the Integrated Pearson distribution
 $\mbox{IP}(\mu;\delta,\beta,\gamma)$,
 $X\sim \mbox{IP}(\mu;\delta,\beta,\gamma)$,
 if there exists a
 quadratic
 $q(x)=\delta x^2+\beta x+\gamma$
 (with $\beta,\delta,\gamma\in\R$, $|\delta|+|\beta|+|\gamma|>0$)
 such that
 \be
 \label{qua1}
 \int_{-\infty}^{x}(\mu-t)f(t)dt=q(x)f(x)
 \ \ \text{for all}  \
 x\in\R.
 \ee
 \end{DEFI}

 \begin{DEFI}
 {\small\bf (Cumulative Ord Family).}
 Let $X$ be an integer-valued r.v.\ with finite mean $\mu$ and
 assume that $p(k)=\Pr(X=k)$, $k\in\Z$,
 is the probability mass function (p.m.f.)\ of $X$.
 We say that $X$ follows the Cumulative Ord distribution
 $\mbox{CO}(\mu;\delta,\beta,\gamma)$,
 $X\sim\mbox{CO}(\mu;\delta,\beta,\gamma)$,
 if there exists a quadratic
 $q(j)=\delta j^2+\beta j+\gamma$
 (with $\beta,\delta,\gamma\in\R$, $|\delta|+|\beta|+|\gamma|>0$)
 such that
 \be
 \label{qua2}
 \sum_{k\le{j}}(\mu-k)p(k)=q(j)p(j)
 \ \ \text{for all}  \
 j\in\Z.
 \ee
 \end{DEFI}

 %\noindent
 It is well known that the commonly used
 distributions are members of the above families,
 e.g.\ Normal, Gamma, Beta, $F$ and $t$ distributions belong to
 the Integrated Pearson
 family, while
 Poisson, Binomial, Pascal (Negative Binomial) and Hypergeometric
 distribution
 are members
 of the Cumulative Ord family.
 Therefore,
 the results of the present note also improve and unify the corresponding bounds for Beta r.v.'s, given by Prakasa Rao (2006) and Wei and Zhang (2009).

 \section{Matrix variance inequalities of Poincar\'{e} and Bessel type}
 \subsection{Continuous Case}
 In this subsection we shall make use
 of the following
 notations.
 \smallskip
 %define the family of functions $\CD^n(X)$, as follows.
 %\smallskip
%
 %\begin{DEFI}
%
 %\noindent
 %{\small\bf Definitions:}
 %{\small\bf (a) Continuous Case.}
 %{\bf Continuous Case.}

 Assume that
 $X\sim\mbox{IP}(\mu;\delta,\beta,\gamma)$
 and 
 denote by $q(x)=\delta x^2+\beta x+\gamma$ its
 quadratic.
 It is known that, under (\ref{qua1}),
 the support $J=J(X)=\{x\in\R:f(x)>0\}$ 
 is a (finite or infinite)
 open interval, say $J(X)=(\alpha,\omega)$ ---
 see \cite{APP1}, \cite{APP2}.
 \smallskip

 %\noindent
 For a fixed integer $n\in\{1,2,\ldots\}$
 we shall denote by
 $\CH^n(X)$ the
 class of functions
 $g:(\alpha,\omega)\to\R$ satisfying the following properties:
 \smallskip

 %\noindent {\bf (a)} $\Var[g(X)]<\infty$.

 \noindent
 H$_1$:
 For each $k\in\{0,1,\ldots,n-1\}$,
 $g^{(k)}$ (with $g^{(0)}=g$)
 is an absolutely continuous function with
 derivative $g^{(k+1)}$.
 \smallskip

 \noindent
 H$_2$:
 For each $k\in\{0,1,\ldots,n\}$,
 $\E[q^k(X) (g^{(k)}(X))^2]<\infty$.
 \bigskip

 %\noindent
 Also, for a fixed integer $n\in\{1,2,\ldots\}$
 we shall denote by
 $\CB^n(X)$ the
 class of functions
 $g:(\alpha,\omega)\to\R$ satisfying the
 following properties:
 \smallskip

 \noindent
 B$_1$: $\Var[g(X)]<\infty$.
 \smallskip

 \noindent
 B$_2$:
 For each $k\in\{0,1,\ldots,n-1\}$,
 $g^{(k)}$ (with $g^{(0)}=g$)
 is an absolutely continuous function with
 derivative $g^{(k+1)}$.
 \smallskip

 \noindent
 B$_3$:
 For each $k\in\{0,1,\ldots,n\}$,
 $\E[q^k(X) |g^{(k)}(X)|]<\infty$.
 \bigskip

 \noindent
 Clearly, if $\E|X|^{2n}<\infty$ then $\CH^n(X)\subseteq
 \CB^n(X)$ (note that H$_2$ (with $k=0$) yields B$_1$), since
 by the C-S inequality,
 $\E^2[q^k(X)|g^{(k)}(X)|]\leq
 \E[q^k(X)]\cdot\E[q^k(X)(g^{(k)}(X))^2]$.
 \vspace{.4em}

 %Here, $g^{(0)}=g$.
 %\vspace{-.3ex}
 %\begin{eqnarray*}
 %\CD^n(X)=\{g:J(X)\vell\R \ |
 %\ g^{(k)} \ \text{is absolutely continuous} \\
 %\text{with derivative $g^{(k+1)}$, $k=0,\ldots,n$}\},
 %%\vspace{-.3ex}
 %\end{eqnarray*}
 %where $g^{(0)}=g$.
 %for a number $n\in\N$
 %\vspace{.5ex}

 Consider now any $p$ functions
 $g_1,\ldots,g_p\in \CH^n(X)$ and set
 $\bbb{g}=(g_1,g_2,\ldots,g_p)^{\mbox{t}}$.
 Then,
 the following $p\times p$
 matrices $\bbb{H}_k=\bbb{H}_k(\bbb{g})$
 are well-defined for $k=1,2,\ldots,n$:
 \be
 \label{c-m eq H}
 {\bbb H}_k=(h_{ij;k}), \ \
 \mbox{where} \ \
 h_{ij;k}:=\E[q^{k}(X)g_i^{(k)}(X)g_j^{(k)}(X)],
 \ \ i,j=1,2,\ldots,p.
 \ee
 Similarly,
 for any functions
 $g_1,\ldots,g_p\in \CB^n(X)$,
 the following $p\times p$
 matrices $\bbb{B}_k=\bbb{B}_k(\bbb{g})$
 are well-defined for $k=1,2,\ldots,n$:
 \be
 \label{c-m eq B}
 {\bbb B}_k=(b_{ij;k}), \ \
 \mbox{where} \ \
 b_{ij;k}:=\E[q^{k}(X)g_i^{(k)}(X)]\cdot
 \E[q^{k}(X)g_j^{(k)}(X)],
 \ \ i,j=1,2,\ldots,p.
 \ee

 Our first result concerns a class of Poincar\'{e}-type
 matrix variance bounds, as follows:
 \begin{theo}
 \label{th.c.Poincare}
 {\rm
 Let $X\sim\IP$ and assume that $\E|X|^{2n}<\infty$ for some
 fixed integer $n\in\{1,2,\ldots\}$.
 Let $g_1,\dots,g_p$ be arbitrary functions in $\CH^{n}(X)$
 and denote by $\bbb{D}=\bbb{D}(\bbb{g})$
 the dispersion matrix of the random
 vector $\bbb{g}=\bbb{g}(X)=(g_1(X),\ldots,g_p(X))^{\mbox{t}}$.
 Also, denote by $\bbb{S}_n=\bbb{S}_n(\bbb{g})$ the
 $p\times p$ matrix
 \[
 \bbb{S}_{n}=\sum_{k=1}^n
 \frac{(-1)^{k-1}}{k!\prod_{j=0}^{k-1}(1-j\delta)}\cdot
 \bbb{H}_k,
 %=\bbb{H}_1-\frac{1}{2! (1-\delta)}\cdot \bbb{H}_2+\cdots+\frac{(-1)^{n-1}}{n!(1-\delta)\cdots(1-(n-1)\delta)}\cdot \bbb{H}_n,
 \]
 where the  matrices $\bbb{H}_k$, $k=1,2,\ldots,n$, are
 defined by (\ref{c-m eq H}).
 Then, the matrix
 \[
 \bbb{A}_n=(-1)^n (\bbb{D}-\bbb{S}_n)
 \]
 is nonnegative definite. Moreover,
 $\bbb{A}_n$ is positive definite unless
 there exist constants $c_1,\ldots,c_p\in\R$, not all zero,
 such that the function $c_1g_1(x)+\cdots+c_pg_p(x)$ is a
 polynomial (in $x$) of degree at most $n$.
 }
 \end{theo}
 \begin{pr}{Proof}
 Fix $\bbb{c}=(c_1,\ldots,c_p)^{\mbox{t}}\in\R^p$ and define the function
 $h_{\bbb{c}}(x)=\bbb{c}^{\mbox{t}} \cdot \bbb{g}(x)=c_1g_1(x)+\cdots+c_pg_p(x)$.
 Since $g_1,\ldots,g_p\in \CH^n(X)$ we see that the function
 $h_{\bbb{c}}$ belongs to $\CH^n(X)$  and, in particular,
 $h_{\bbb{c}}(X)$
 has finite variance
 %$h_{\bbb{c}}\in\CD^m(X)$ for all $m=0,1,\ldots,n$
 and
 $\E[q^{k}(X)(h^{(k)}_{\bbb{c}}(X))^2]<\infty$,
 $k=1,2\ldots,n$.
 Thus, we can make use of the inequality
 (see \cite{APP1}, \cite{Joh}; c.f.\ \cite{HK},
 \cite{Pap})
 \[
 (-1)^n[\Var h_{\bbb{c}}(X)-S_n]\ge0, \ \
 \mbox{where} \ \
 S_n=
 \sum_{k=1}^{n}\frac{(-1)^{k-1}}{k!\prod_{j=0}^{k-1}(1-j\delta)}
 \E[q^k(X)(h_{\bbb{c}}^{(k)}(X))^2],
 \]
 in which the equality holds if and only if
 $h_{\bbb{c}}$ is a polynomial of degree at most $n$.
 It is well-known that $\Var h_{\bbb{c}}(X)=\bbb{c}^{\mbox{t}} \bbb{D}\bbb{c}$ and
 it is easily seen that
 $\E[q^k(X)(h_{\bbb{c}}^{(k)}(X))^2]=\bbb{c}^{\mbox{t}} \bbb{H}_k
 \bbb{c}$ with $\bbb{H}_k$ ($k=1,\ldots,n$) as in
 (\ref{c-m eq H}). Thus,
 $S_n=\bbb{c}^{\mbox{t}} {\bbb{S}}_n \bbb{c}$ and the preceding inequality
 takes the form
 \[
 \bbb{c}^{\mbox{t}} \left[(-1)^n (\bbb{D}-\bbb{S}_n)
 \right]\bbb{c}\geq 0.
 \]
 Since $\bbb{c}\in\R^p$ is arbitrary it follows
 that the matrix $(-1)^n (\bbb{D}-\bbb{S}_n)$ is
 nonnegative definite. Clearly
 the inequality is strict for all $\bbb{c}\in \R^p$
 for which $h_{\bbb{c}}\notin\spam[1,x,\ldots,x^{n}]$.
 $\Box$
 \end{pr}
 \bigskip

 \noindent
 {\small\bf Remark 1.}  
 (a)
 Olkin and Shepp's (2005) matrix inequalities are particular
 cases of Theorem \ref{th.c.Poincare} for $n=1$  and with
 $X$ being a standard
 Normal or a Gamma r.v. For example, when $n=1$ and
 $X=Z\sim N(0,1)\equiv\mbox{IP}(0;0,0,1)$
 then
 $\bbb{S}_1=\bbb{H}_1=\bbb{H}=(\E[g_i'(Z)g_j'(Z)])_{p\times p}$
 and we get the inequality $\bbb{D}\leq \bbb{H}$ (in the
 Loewner ordering). Moreover, for $X=Z$ and $n=2$ or $3$, Theorem
 \ref{th.c.Poincare} yields the new  matrix variance bounds
 $\bbb{H}-\frac{1}{2}\bbb{H}_2\leq \bbb{D}$ and
 $\bbb{D}\leq \bbb{H}-\frac{1}{2}\bbb{H}_2+\frac{1}{6}\bbb{H}_3 $
 where
 $\bbb{H}_2=(\E[g''_i(Z)g''_j(Z)])_{p\times p}$ and
 $\bbb{H}_3=(\E[g'''_i(Z)g'''_j(Z)])_{p\times p}$.
 \vspace{.3em}

 \noindent
 (b)
 Theorem 1 applies to Beta r.v.'s. In particular,  when $n=1$ and $X\sim$Beta$(a,b)$ (with p.d.f.\
 $f(x)\propto x^{a-1}(1-x)^{b-1}$, $0<x<1$, and parameters $a,b>0$) then
 $q(x)=x(1-x)/(a+b)$. Theorem 1 yields the inequality $\bbb{D}\leq \bbb{H}$ (in the Loewner ordering) where
 $\bbb{H}=\frac{1}{a+b}
 (\E[X(1-X)g'_i(X)g'_j(X)])_{p\times p}$.
 This compares with Theorem 4.1 in
 Prakasa Rao (2006); c.f.\
 Wei and Zhang (2009), Remark 1.
 %\bigskip
 \vspace{1em}

 Next we show some similar Bessel-type
 matrix variance bounds. The particular case  
 of a Beta r.v.\ is covered by Theorem 1 of Wei 
 and Zhang (2009). 
 \begin{theo}
 \label{th.c.Bessel}
 {\rm
 Let $X\sim\IP$ and assume that $\E|X|^{2n}<\infty$ for some
 fixed integer $n\in\{1,2,\ldots\}$.
 Let $g_1,\dots,g_p$ be arbitrary functions in $\CB^{n}(X)$
 and denote by $\bbb{D}=\bbb{D}(\bbb{g})$
 the dispersion matrix of the random
 vector $\bbb{g}=\bbb{g}(X)=(g_1(X),\ldots,g_p(X))^{\mbox{t}}$.
 Also, denote by $\bbb{L}_n=\bbb{L}_n(\bbb{g})$ the
 $p\times p$ matrix
 \[
 \bbb{L}_{n}=\sum_{k=1}^n
 \frac{1}{k! \ \E[q^k(X)]\prod_{j=k-1}^{2k-2}(1-j\delta)}\cdot
 \bbb{B}_k,
 \]
 where the matrices $\bbb{B}_k$, $k=1,2,\ldots,n$, are
 defined by (\ref{c-m eq B}).
 Then,
 \[
  \bbb{L}_n \leq\bbb{D}
 \]
 in the Loewner ordering. Moreover,
 $\bbb{D}-\bbb{L}_n$ is positive definite unless
 there exist constants $c_1,\ldots,c_p\in\R$, not all zero,
 such that the function $c_1g_1(x)+\cdots+c_pg_p(x)$ is a
 polynomial (in $x$) of degree at most $n$.
 }
 \end{theo}
 \begin{pr}{Proof}
 Fix $\bbb{c}=(c_1,\ldots,c_p)^{\mbox{t}}\in\R^p$ and, as in the previous proof, define the function
 $h_{\bbb{c}}(x)=\bbb{c}^{\mbox{t}} \cdot \bbb{g}(x)=c_1g_1(x)+\cdots+c_pg_p(x)$.
 Since $g_1,\ldots,g_p\in \CB^n(X)$ we see that the function
 $h_{\bbb{c}}$ belongs to $\CB^n(X)$. In particular,
 $h_{\bbb{c}}(X)$
 has finite variance
 %$h_{\bbb{c}}\in\CD^m(X)$ for all $m=0,1,\ldots,n$
 and
 $\E[q^{k}(X)|h^{(k)}_{\bbb{c}}(X)|]<\infty$,
 $k=1,2\ldots,n$.
 Thus, we can apply the inequality
 %(see \cite{APP1}, \cite{Joh}; c.f.\ \cite{HK},
 %\cite{Pap})
 (see \cite{APP2})
 \[
 \Var h_{\bbb{c}}(X)\geq L_n, \ \
 \mbox{where} \ \
 L_n=
 \sum_{k=1}^{n}\frac{\E^2[q^k(X) h_{\bbb{c}}^{(k)}(X)]}{k!
 \  \E[q^k(X)] \prod_{j=k-1}^{2k-2}(1-j\delta)},
 \]
 in which the equality holds if and only if
 $h_{\bbb{c}}$ is a polynomial of degree at most $n$.
 Observe that $\Var h_{\bbb{c}}(X)=\bbb{c}^{\mbox{t}} \bbb{D}\bbb{c}$ and
 $\E^2[q^k(X)|h_{\bbb{c}}^{(k)}(X)|]=\bbb{c}^t \bbb{B}_k
 \bbb{c}$ with $\bbb{B}_k$ ($k=1,\ldots,n$) as in
 (\ref{c-m eq B}). Thus,
 $L_n=\bbb{c}^{\mbox{t}} {\bbb{L}}_n \bbb{c}$ and the preceding inequality
 takes the form
 \[
 \bbb{c}^{\mbox{t}} \left[\bbb{D}-\bbb{L}_n
 \right]\bbb{c}\geq 0.
 \]
 Since $\bbb{c}\in\R^p$ is arbitrary it follows
 that the matrix $\bbb{D}-\bbb{L}_n$ is
 nonnegative definite. Clearly
 the inequality is strict for all $\bbb{c}\in \R^p$
 for which $h_{\bbb{c}}\notin\spam[1,x,\ldots,x^{n}]$.
 $\Box$
 \end{pr}
 %\bigskip

 \subsection{Discrete Case}
 In this subsection we shall make use of the following
 notations.
 \smallskip
 %define the family of functions $\CD^n(X)$, as follows.
 %\smallskip
%
 %\begin{DEFI}
%
 %\noindent
 %{\small\bf Definitions:}
 %{\small\bf (a) Continuous Case.}
 %{\bf Continuous Case.}

 Assume that
 $X\sim\mbox{CO}(\mu;\delta,\beta,\gamma)$.
 It is known (see \cite{APP1}, \cite{APP2}) that, 
 under {\rm(\ref{qua2})}, the
 support $J=J(X)=\{k\in\Z:p(k)>0\}$ is a (finite of infinite) interval of integers, say
 $J(X)=\{\alpha,\alpha+1,\ldots,\omega-1,\omega\}$.
 Write $q(x)=\delta x^2+\beta x+\gamma$ for the quadratic of $X$ and let $q^{[k]}(x)=q(x)q(x+1)\cdots q(x+k-1)$
 for $k=1,2,\ldots$ (with $q^{[0]}(x)\equiv 1$, $q^{[1]}(x)\equiv
 q(x)$). For any function $g:\Z\to\R$ we shall denote by
 $\Delta^k[g(x)]$ its $k$-th forward difference, i.e.,
 $\Delta^k[g(x)]=\Delta[\Delta^{k-1}[g(x)]]$,
 $k=1,2,\ldots$,
 with $\Delta[g(x)]=g(x+1)-g(x)$ and
 $\Delta^0[g(x)]\equiv  g(x)$.
 \smallskip
 %
 
 %\noindent
 For a fixed integer $n\in\{1,2,\ldots\}$
 we shall denote by
 $\CH_d^n(X)$ the
 class of functions
 $g:J(X)\to\R$ satisfying the following property:
 \smallskip

 %\noindent {\bf (a)} $\Var[g(X)]<\infty$.

 \noindent
 HD$_1$:
 For each $k\in\{0,1,\ldots,n\}$,
 $\E[q^{[k]}(X) (\Delta^k[g(X)])^2]<\infty$.
 \smallskip

 %\noindent
 Also, for a fixed integer $n\in\{1,2,\ldots\}$
 we shall denote by
 $\CB_d^n(X)$ the
 class of functions
 $g:J(X)\to\R$ satisfying the
 following properties:
 \smallskip

 \noindent
 BD$_1$: $\Var[g(X)]<\infty$.
 \smallskip

 \noindent
 BD$_2$:
 For each $k\in\{0,1,\ldots,n\}$,
 $\E[q^k(X) |\Delta^k[g(X)]|]<\infty$.
 \smallskip

 \noindent
 Clearly, if $\E|X|^{2n}<\infty$ then $\CH_d^n(X)\subseteq
 \CB_d^n(X)$ (note that HD$_1$ (with $k=0$) yields BD$_1$).
 Indeed, since $\Pr[q^{[k]}(X)\geq 0]=1$,
 the C-S inequality implies that $\E^2[q^{[k]}(X)|\Delta^k[g(X)]|]\leq
 \E[q^{[k]}(X)]\cdot\E[q^{[k]}(X)(\Delta^k[g(X)])^2]$.
 \vspace{.4em}

 %Here, $g^{(0)}=g$.
 %\vspace{-.3ex}
 %\begin{eqnarray*}
 %\CD^n(X)=\{g:J(X)\vell\R \ |
 %\ g^{(k)} \ \text{is absolutely continuous} \\
 %\text{with derivative $g^{(k+1)}$, $k=0,\ldots,n$}\},
 %%\vspace{-.3ex}
 %\end{eqnarray*}
 %where $g^{(0)}=g$.
 %for a number $n\in\N$
 %\vspace{.5ex}

 Consider now any $p$ functions
 $g_1,\ldots,g_p\in \CH_d^n(X)$ and set
 $\bbb{g}=(g_1,g_2,\ldots,g_p)^{\mbox{t}}$.
 Then,
 the following $p\times p$
 matrices $\bbb{H}_k=\bbb{H}_k(\bbb{g})$
 are well-defined for $k=1,2,\ldots,n$:
 \be
 \label{c-m eq Hd}
 {\bbb H}_k=(h_{ij;k}), \ \
 \mbox{where} \ \
 h_{ij;k}:=\E[q^{[k]}(X)\Delta^k [g_i(X)] \Delta^k [g_j(X)]],
 \ \ i,j=1,2,\ldots,p.
 \ee
 Similarly,
 for any functions
 $g_1,\ldots,g_p\in \CB_d^n(X)$,
 the following $p\times p$
 matrices $\bbb{B}_k=\bbb{B}_k(\bbb{g})$
 are well-defined for $k=1,2,\ldots,n$:
 \begin{eqnarray}
 %\label{c-m eq Bd}
 && {\bbb B}_k=(b_{ij;k}),
 \ \
 \mbox{where}
 \nonumber
 \\
 \label{c-m eq Bd}
 &&
 b_{ij;k}:=\E[q^{[k]}(X)\Delta^k[g_i(X)]]\cdot
 \E[q^{[k]}(X)\Delta^k[g_j(X)]],
 \ \ i,j=1,2,\ldots,p.
 \end{eqnarray}

 The matrix variance inequalities for the discrete case
 are summarized in the
 following theorem; its proof, being the same as in the
 continuous case, is omitted.
 \begin{theo}
 \label{th.d.Poincare.Bessel}
 {\rm
 Let $X\sim\CO$ and assume that $\E|X|^{2n}<\infty$ for some
 fixed integer $n\in\{1,2,\ldots\}$.
 \vspace{.2em}

 \noindent
 (a)
 Let $g_1,\dots,g_p$ be arbitrary functions in $\CH_d^{n}(X)$
 and denote by $\bbb{D}=\bbb{D}(\bbb{g})$
 the dispersion matrix of the random
 vector $\bbb{g}=\bbb{g}(X)=(g_1(X),\ldots,g_p(X))^{\mbox{t}}$.
 Also, denote by $\bbb{S}_n=\bbb{S}_n(\bbb{g})$ the
 $p\times p$ matrix
 \[
 \bbb{S}_{n}=\sum_{k=1}^n
 \frac{(-1)^{k-1}}{k!\prod_{j=0}^{k-1}(1-j\delta)}\cdot
 \bbb{H}_k,
 \]
 where the  matrices $\bbb{H}_k$, $k=1,2,\ldots,n$, are
 defined by (\ref{c-m eq Hd}).
 Then, the matrix
 \[
 \bbb{A}_n=(-1)^n (\bbb{D}-\bbb{S}_n)
 \]
 is nonnegative definite. Moreover,
 $\bbb{A}_n$ is positive definite unless
 there exist constants $c_1,\ldots,c_p\in\R$, not all zero,
 and a polynomial $P_n:\R\to\R$, of degree at most $n$,
 such that $\Pr[c_1g_1(X)+\cdots+c_pg_p(X)=P_n(X)]=1$.
 \vspace{.2em}

 \noindent
 (b)
 Let $g_1,\dots,g_p$ be arbitrary functions in $\CB_d^{n}(X)$
 and denote by $\bbb{D}=\bbb{D}(\bbb{g})$
 the dispersion matrix of the random
 vector $\bbb{g}=\bbb{g}(X)=(g_1(X),\ldots,g_p(X))^{\mbox{t}}$.
 Also, denote by $\bbb{L}_n=\bbb{L}_n(\bbb{g})$ the
 $p\times p$ matrix
 \[
 \bbb{L}_{n}=\sum_{k=1}^n
 \frac{1}{k! \ \E[q^{[k]}(X)]\prod_{j=k-1}^{2k-2}(1-j\delta)}\cdot
 \bbb{B}_k,
 \]
 where the matrices $\bbb{B}_k$, $k=1,2,\ldots,n$, are
 defined by (\ref{c-m eq Bd}).
 Then,
 \[
  \bbb{L}_n \leq\bbb{D}
 \]
 in the Loewner ordering. Moreover,
 $\bbb{D}-\bbb{L}_n$ is positive definite
 unless
 there exist constants $c_1,\ldots,c_p\in\R$, not all zero,
 and a polynomial $P_n:\R\to\R$, of degree at most $n$,
 such that $\Pr[c_1g_1(X)+\cdots+c_pg_p(X)=P_n(X)]=1$.
 %unless
 %there exist constants $c_1,\ldots,c_p\in\R$, not all zero,
 %such that the function $c_1g_1(x)+\cdots+c_pg_p(x)$ is a
 %polynomial of degree at most $n$.
 \newline
 [It should be noted that the $k$-th term in the sum
 defining the matrix $\bbb{S}_n$ or the matrix $\bbb{L}_n$,
 above, should be treated as the null matrix,
 $\bbb{0}_{p\times p}$, whenever $\E[q^{[k]}(X)]=0$.]
 }
 \end{theo}

 As an example consider the case where $X\sim\mbox{Poisson}(\lambda)$
 with p.m.f.\ $p(k)=e^{-\lambda}\lambda^k/k!$,
 $k=0,1,\ldots$ ($\lambda>0$). Then
 $X\sim\mbox{CO}(\lambda;0,0,\lambda)$ so that $q(x)\equiv
 \lambda$. It follows that
 $\bbb{H}_k=\lambda^k (\E[\Delta^k[g_i(X)]\Delta^k[g_j(X)]])_{p\times p}$
 and
 $\bbb{B}_k=\lambda^{2k} (\E[\Delta^k[g_i(X)]]\cdot\E[\Delta^k[g_j(X)]])_{p\times
 p}$. Thus, for $n=1$ and $p=2$ Theorem
 \ref{th.d.Poincare.Bessel}(a) yields the matrix inequality
 \[
 \left(
 \begin{array}{cc}
 \Var[g_1] & \Cov[g_1,g_2] \\
 \Cov[g_1,g_2] & \Var[g_2]
 \end{array}
 \right)
 \leq
 \lambda
 \left(
 \begin{array}{cc}
 \E[(\Delta[g_1])^2] & \E[\Delta[g_1]\Delta[g_2]] \\
 \E[\Delta[g_1]\Delta[g_2]] & \E[(\Delta[g_2])^2]
 \end{array}
 \right).
 \]

\end{document}